\documentclass[preprint,preprintnumbers,amsmath,amssymb]{revtex4}
\usepackage[abs]{overpic}
\usepackage{graphicx,axodraw,subfigure}
\begin{document}
\preprint{HUPD1209}
\def\sla#1{\rlap/#1}
\def\tbr{\textcolor{red}}
\def\tcr{\textcolor{red}}
\def\ov{\overline}
\def\dprime{{\prime \prime}}
\def\nn{\nonumber}
\def\f{\frac}
\def\beq{\begin{equation}}
\def\eeq{\end{equation}}
\def\bea{\begin{eqnarray}}
\def\eea{\end{eqnarray}}
\def\bsub{\begin{subequations}}
\def\esub{\end{subequations}}
\def\dc{\stackrel{\leftrightarrow}{\partial}}
\def\ynu{y_{\nu}}
\def\ydu{y_{\triangle}}
\def\ynut{{y_{\nu}}^T}
\def\ynuv{y_{\nu}\frac{v}{\sqrt{2}}}
\def\ynuvt{{\ynut}\frac{v}{\sqrt{2}}}
\def\d{\partial}
\def\nn{\nonumber}
\def\beq{\begin{equation}}
\def\eeq{\end{equation}}
\def\bei{\begin{itemize}}
\def\eei{\end{itemize}}
\def\bea{\begin{eqnarray}}
\def\eea{\end{eqnarray}}
\def\ynu{y_{\nu}}
\def\ydu{y_{\triangle}}
\def\ynut{{y_{\nu}}^T}
\def\ynuv{y_{\nu}\frac{v}{\sqrt{2}}}
\def\ynuvt{{\ynut}\frac{v}{\sqrt{2}}}
\def\s{\partial \hspace{-.47em}/}
\def\ad{\overleftrightarrow{\partial}}
\def\ss{s \hspace{-.47em}/}
\def\pp{p \hspace{-.47em}/}
\def\bos{\boldsymbol}
\title{The form factors for hadronic tau decay using\\
background field method}

\author{Daiji Kimura$^a$, Kang Young Lee$^b$, Takuya Morozumi$^c$}

\address{$^a$ Faculty of Engineering, Kinki University,
Takaya, Higashi-Hiroshima, Japan 739-2116 \\
$^b$ Department of Physics Education and  Education Research Institute,
Gyeongsang National University, Jinju 660-701, Korea \\
$^c$ Graduate School of Science, Hiroshima University,
1-3-1 Kagami-Yama, Higashi-Hiroshima, Japan 739-8526}
\begin{abstract}
In the two body hadronic tau decays, such as $\tau \to K \pi(\eta) \nu$, 
vector mesons play important role. Belle and Babar
measured hadronic invariant mass spectrum of $\tau \to K \pi \nu$ decay.
To compare the spectrum with theoretical
prediction,  we develop the chiral Lagrangian with vector mesons in in \cite{Kimura:2012nx}. 
We compute the form factors of the hadronic 
$\tau$ decay taking account of the quantum corrections of Nambu Goldstone bosons.
We also show how to renormalize
the divergence of the Feynman diagrams with arbitrary number of loops and determine the
counterterms within one loop using background field method.
In this report, we discuss the renormalization of \cite{Kimura:2012nx} 
by considering the one loop Feynman diagrams.
\end{abstract}
\maketitle





\section{Introduction}
In tau lepton decay into two pseudoscalar mesons, 
the vector form factor is an important quantity.
Since the decay of tau lepton releases very energetic hadrons
$E_h \sim 1.7$ (GeV), one must include  vector mesons into the 
effective theory. 
Our aim is systematic and quantitative study of the form factors.
In \cite{Kimura:2012nx}, we establish the systematic renormalization program for the chiral
Lagrangian including the vector mesons by determining the possible 
form of the counterterms for diagrams with arbitrary number of loops of Nambu-Goldstone
boson. 
\section{Leading order chiral Lagrangian with vector meson}
In what follows, we ignore the chiral breaking terms and $\eta^\prime$
meson for simplicity.
We start with the following chiral Lagrangian with vector mesons.
\bea
{\cal L}=\frac{f^2}{4} {\rm Tr} (D_{L\mu} U {D_L}^\mu U^{\dagger})
+M_{V}^2 {\rm Tr}(V_{\mu}-\frac{\alpha_{\mu}}{g})^2,  
\label{eq:lag} 
\eea
where $U=\xi^2$ with $\xi=e^{i \frac{\pi}{f}}$.
$D_{L}$ and $\alpha_\mu$ are given by,
\bea
D_{L\mu} U &=& (\partial_\mu + iA_{L\mu})U , \nn \\
\alpha_{\mu}&=&\frac{\xi^\dagger D_{L \mu} \xi+\xi \partial_\mu \xi^\dagger}
{2i}.
\eea
$A_L$ denotes the external gauge fields.
\section{The superficial degree of divergence and renormalization}
In \cite{Kimura:2012nx}, we show the degree of the divergence
$\omega$ for the 1 particle irreducible (1 PI)
diagram with $N$ Nambu-Goldstone boson loops and $N_V$ external vector meson 
legs is given as,
\bea
\omega=2N+2-N_V.
\label{eq:omega}
\eea
Below we count the degree of divergence
in one loop Feynman diagrams explicitly and compare them with the formulae 
in Eq.(\ref{eq:omega}).
\begin{itemize}
\item The first example is a self-energy diagram of vector meson
shown in Fig.1. 
Note that $V \to 2 \pi$ interaction vertex in the lowest order is proportional to, 
\bea
{\rm Tr}(V^\mu [\pi, \partial_\mu \pi]). 
\eea
The superficial divergence $\omega$ of Fig.1
is given as,
\bea
p^\omega&=& p^4 \left(\frac{1}{p^2}\right)^2 p^2 =p^2,
\eea  
which leads to $\omega=2$.
\item The next example is $V \to 2 \pi$ vertex in Fig.2.
In this case, the index of the divergence is $3$, since,
\bea
p^\omega=p^4 p p^2 \left(\frac{1}{p^2}\right)^2=p^3.
\eea
\item
Finally, one can compute the degree of the divergence
for the four vector meson vertex 
which corresponds to Fig.3.
\bea
p^\omega=p^4 \left(\frac{1}{p^2}\right)^4 p^4 =p^0.
\eea
Therefore the index of the divergence is $\omega=0$ and it implies
the logarithmic divergence.
\end{itemize}
\begin{figure}[htbp]
\begin{center}
\includegraphics[width=7cm]{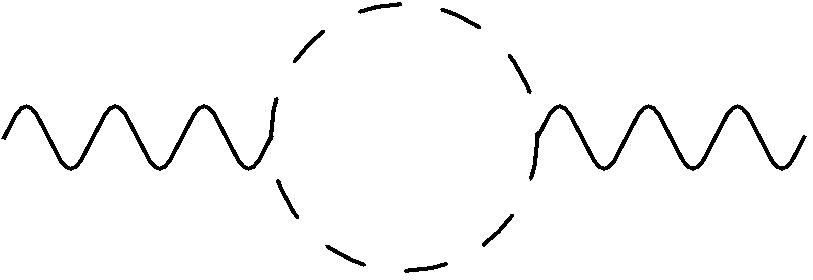}
\caption{The self-energy diagram for vector meson}
\end{center}
\end{figure}
\begin{figure}[htbp]
\begin{center}
\includegraphics[width=7cm]{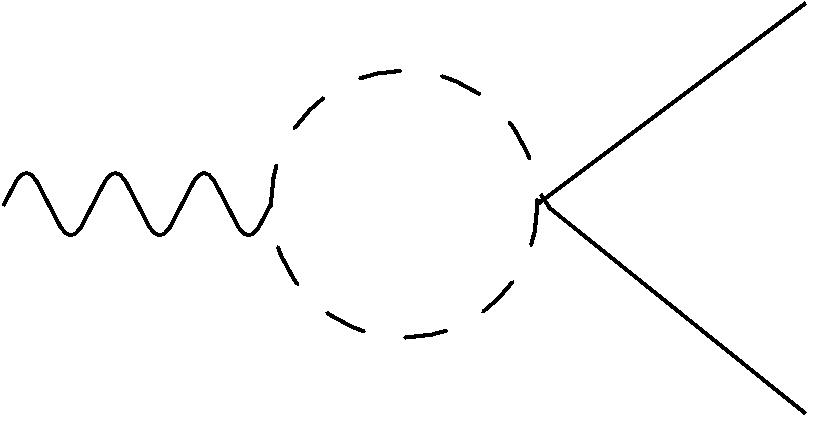}
\end{center}
\caption{One loop correction to $V \to PP$ vertex.}
\end{figure}
\begin{figure}[htbp]
\begin{center}
\includegraphics[width=5cm]{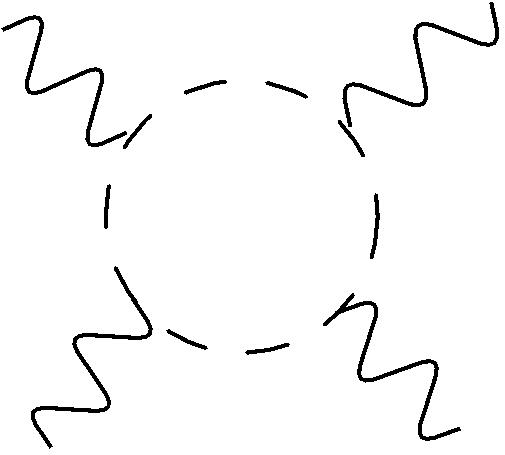}
\end{center}
\caption{four vector mesons vertex}
\end{figure}
The indices for three cases in the above coincide with the Eq.(\ref{eq:omega})
for $N=1$ and $N_V=2,1,4$ respectively.

In \cite{Kimura:2012nx}, we compute the Nambu Goldstone boson one loop corrections and 
determine the counterterms.
To compute the loop correction, we use the background field method. With the method,
one can include loop corrections consistent with the chiral symmetry. The chiral invariant part 
of the counterterms is given as,
\def\aTr{{\rm Tr}}
\bea
 L_c&=&
K_1  2 i \aTr(\alpha_{\perp \mu} \alpha_{\perp \nu})
(D_\mu v_\nu-D_\nu v_\mu+i[v_\mu, v_\nu])\nn \\
&-&
\frac{1}{2}K_2
\aTr(\xi^\dagger F_{L \mu \nu} \xi)(D_\mu v_\nu-D_\nu v_\mu+i[v_\mu, v_\nu])\nn \\
&-&\frac{1}{2}
K_3 \aTr(D_\mu v_\nu-D_\nu v_\mu+i[v_\mu, v_\nu])^2
\nn \\
&+& K_6 \aTr(v_\rho \alpha_\perp^\mu)  \aTr(v^\rho \alpha_{\perp \mu})+ 
K_7\aTr(v^2 \alpha_{\perp \mu} \alpha_{\perp}^\mu)\nn \\
&+&K_8
\aTr(v^2) \aTr(\alpha_{\perp \mu} \alpha_\perp^\mu) \nn \\
&+&K_9\{\aTr(v^2)\}^2+K_{10} \aTr(v^4) \nn \\  
&+&L_1 \{\aTr(D_{L\mu}U(D_{L}^\mu U)^\dagger)\}^2 \nn \\
&+&L_2  \aTr \{D_{L}^\mu U (D_L^\nu U)^\dagger\} 
\aTr \{D_{L\mu}U (D_{L \nu} U)^\dagger\} \nn \\
&+&L_3 \aTr \{D_{L}^\mu U (D_{L \mu} U)^\dagger D_{L}^\nu U 
(D_{L \nu} U)^\dagger\} \nn \\
&+& i L_9 
\aTr\{F_{L \mu \nu} D^{\mu} U (D^{\nu} U)^\dagger\} \nn \\
&+& H_1 \aTr F_{L \mu \nu} F_L^{\mu \nu}.
\label{eq:fullcounter}
\eea
where
\bea
v_\mu&=&\frac{M_V^2}{2 gf^2}(V_\mu -\frac{\alpha_\mu}{g}),\nn \\
D_\mu v_\nu&=&\partial_\mu v_\nu + i [\alpha_\mu, v_\nu], \nn \\
\alpha_{\perp \mu}&=&\frac{\xi^\dagger D_{L \mu} \xi-\xi \partial_\mu \xi^\dagger}
{2i}.
\eea
One can easily see $K_3$ corresponds to the counterterm for self-energy of vector meson.
As we expect, it contains the second derivatives on the vector mesons. 
$V \to PP$ vertex can be renormalized by the terms proportional to $K_1$ and $K_3$. 
The four vector meson
vertex is renormalized by the terms proportional to $K_9$ and $K_{10}$.
The number of the derivatives in the counterterms of Eq.(\ref{eq:fullcounter}) 
coincides with the superficial degree of the divergence. 

\nocite{*}
\bibliographystyle{elsarticle-num}
\bibliography{martin}



\noindent
{\bf Acknowledgements}\\
K.Y.L.
was supported by the Basic Science Research Program through the NRF funded by MEST
(2010-0010916). T. M. was supported by KAKENHI, Grant-in-Aid for Scientific
Research(C) No.22540283 from JSPS, Japan.

\end{document}